\documentclass{article}

\usepackage{microtype}
\usepackage{graphicx}
\usepackage{subcaption}
\usepackage{booktabs} %

\usepackage{hyperref}

\PassOptionsToPackage{capitalize,noabbrev}{cleveref}

\usepackage[accepted]{icml2026}

\usepackage{amsmath,amsfonts,bm}

\def\eqref#1{equation~\ref{#1}}

\def\1{\bm{1}}

\DeclareMathAlphabet{\mathsfit}{\encodingdefault}{\sfdefault}{m}{sl}
\SetMathAlphabet{\mathsfit}{bold}{\encodingdefault}{\sfdefault}{bx}{n}

\usepackage{booktabs}
\usepackage{array}
\usepackage{xurl}
\usepackage{balance}
\usepackage{xspace}
\usepackage{tikz}
\usepackage{pgfplots}
\usepackage{tabularx}
\usepackage{threeparttable}
\usepackage{pifont}
\usepackage[htt]{hyphenat}
\newcommand{\ourmethod}{{SlotGuard}\xspace}
\newcommand{\codebaseurl}{\url{https://github.com/illinoisdata/SlotGuard}}
\usepackage{enumitem}
\usepackage[scaled=0.88]{beramono}
\let\oldpath\path
\renewcommand{\path}[1]{{\small\oldpath{#1}}}

\usepackage[scaled=0.82]{inconsolata}
\usepackage[T1]{fontenc}
\usepackage{textcomp}
\usepackage{xcolor}
\usepackage[most]{tcolorbox}
\tcbuselibrary{listings}

\newtcblisting{slotlisting}[1][]{
  listing only,
  listing options={
    basicstyle=\ttfamily\footnotesize,
    columns=fullflexible,
    keepspaces=true,
    upquote=true,
    showstringspaces=false,
    breaklines=true,
    escapeinside={(*@}{@*)},
  },
  colback=white,
  colframe=gray,
  boxrule=0.4pt,
  arc=0pt,
  left=2pt,
  right=2pt,
  top=0pt,
  bottom=0pt,
  before skip=0.35em,
  after skip=0.35em,
  title={#1},
  fonttitle=\bfseries,
}

\usepackage{tikz}
\usetikzlibrary{arrows.meta,positioning,fit,backgrounds}

\newcommand{\nop}[1]{}

\newcommand{\sysname}{SlotGuard\xspace}

\usetikzlibrary{arrows.meta,positioning,calc,shapes.geometric,decorations.text,patterns,patterns.meta,fit,backgrounds,chains,shadows,math,circuits.ee.IEC,decorations.pathmorphing, decorations.pathreplacing, decorations.shapes}

\definecolor{colorset1}{RGB}{169, 194, 225}
\definecolor{colorset2}{RGB}{61, 104, 168}
\definecolor{colorset3}{RGB}{31, 60, 104}
\definecolor{colorset4}{RGB}{35, 31, 32}
\definecolor{colorset5}{RGB}{0, 0, 0}         %
\definecolor{colorLazy}{RGB}{247, 168, 0}

\definecolor{barFull}{HTML}{2E64A1}
\definecolor{barLazy}{HTML}{F3B404}
\definecolor{barLightBlue}{HTML}{A9CCE3}

\definecolor{cAmain}{HTML}{1f77b4}  %
\colorlet{cAlight}{cAmain!25}
\definecolor{cBmain}{HTML}{ff7f0e}  %
\colorlet{cBlight}{cBmain!25}
\definecolor{cCmain}{HTML}{2ca02c}  %
\colorlet{cClight}{cCmain!25}
\definecolor{cDmain}{HTML}{d62728}  %
\colorlet{cDlight}{cDmain!25}
\definecolor{cEmain}{HTML}{9467bd}  %
\colorlet{cElight}{cEmain!25}
\definecolor{cFmain}{HTML}{8c564b}  %
\colorlet{cFlight}{cFmain!25}
\definecolor{cGmain}{HTML}{e377c2}  %
\colorlet{cGlight}{cGmain!25}
\definecolor{cZmain}{HTML}{030303}  %
\colorlet{cZlight}{cZmain!25}
\colorlet{cZlightlight}{cZmain!5}
\definecolor{cPositivemain}{HTML}{2ca02c}  %
\colorlet{cPositivelight}{cPositivemain!25}
\definecolor{cNegativemain}{HTML}{d62728}  %
\colorlet{cNegativelight}{cNegativemain!25}

\definecolor{MyRed}{RGB}{228,26,28}   
\definecolor{MyBlue}{RGB}{55,126,184}  
\definecolor{MyGreen}{RGB}{77,175,74}    
\definecolor{MyOrange}{RGB}{255,127,0}  
\definecolor{MyPurple}{RGB}{152,78,163}
\definecolor{MyBrown}{RGB}{166,86,40}   
\definecolor{MyPink}{RGB}{247,129,191} 
\definecolor{MyGray}{RGB}{153,153,153}   

\newcommand*{\circled}[1]{\lower.7ex\hbox{\tikz\draw (0pt, 0pt)%
    circle (.5em) node {\makebox[1em][c]{\small #1}};}}

\usepackage{tikz}
\usetikzlibrary{arrows.meta,positioning,fit,calc}

\usepackage[most]{tcolorbox}
\usepackage{xcolor}

\newtcolorbox{examplebox}{
  colback=gray!4,
  colframe=black!25,
  boxrule=0.4pt,
  arc=1.5mm,
  left=1.5mm,
  right=1.5mm,
  top=1mm,
  bottom=1mm,
  fonttitle=\bfseries,
  title=Example provider-bound transcript
}
\usepackage{cleveref}

\crefformat{section}{§#2#1#3}
\Crefformat{section}{§#2#1#3}
\crefformat{subsection}{§#2#1#3}
\Crefformat{subsection}{§#2#1#3}
\crefformat{appendix}{§#2#1#3}
\Crefformat{appendix}{§#2#1#3}

\usepackage[skip=4pt]{caption}
\usepackage{amsmath}
\usepackage{amssymb}
\usepackage{mathtools}
\usepackage{amsthm}

\theoremstyle{plain}

\theoremstyle{definition}

\theoremstyle{remark}

\usepackage[textsize=tiny]{todonotes}

\icmltitlerunning{SlotGuard: Stop Oversharing Private Local Context in LLM Agent Transcripts}

\begin{document}

\twocolumn[
\icmltitle{\ourmethod: Stop Oversharing Private Local Context in\\ LLM Agent Transcripts}

  \icmlsetsymbol{equal}{*}

  \begin{icmlauthorlist}
    \icmlauthor{Haocheng Xia}{uiuc}
    \icmlauthor{Yongjoo Park}{uiuc}
  \end{icmlauthorlist}

  \icmlaffiliation{uiuc}{Siebel School of Computing and Data Science, University of Illinois Urbana-Champaign}

  \icmlcorrespondingauthor{Yongjoo Park}{yongjoo@illinois.edu}

  \icmlkeywords{Machine Learning, ICML}

  \vskip 0.3in
]

\printAffiliationsAndNotice{}  %

\begin{abstract}
LLM agents can expose private local context (e.g., paths, emails) and credentials (e.g., API keys) as agent observations (e.g., tool outputs, shell logs, and file reads) are appended to provider-bound transcripts. 
Existing placeholder redaction is brittle: it can miss embedded or cross-turn references, over-redact benign lookalikes, and destroy the structure useful for reasoning. We present \ourmethod, a local transcript boundary that can hide sensitive data while retaining agents' performance. \ourmethod rewrites structural bindings as typed, suffix-aware slots, replaces secrets with format-preserving synthetic values, links cross-turn references with a lightweight session graph, and restores raw values only inside the trusted runtime. On controlled repository-oriented agent transcripts, \ourmethod removes all 20,814 annotated structurally sensitive characters across 9,229 paths and reduces credential leakage to 0.0\% across 852 planted values. It remains close to raw-transcript task success across four upstream models, while generic redaction drops to 2.5\%. 
Deterministic transcript rewriting takes a median of 14.424~$\mu$s per agent turn, excluding the optional semantic detector.
The code is publicly accessible at \url{https://github.com/illinoisdata/SlotGuard}.
\end{abstract}

\section{Introduction}

LLM agents access sensitive information.
To perform various tasks (e.g., development, enterprise automation, data analysis),
agents employ various tools to interact with file systems, shells, processes, configuration files, and external services on the user's behalf~\cite{gpt-codex,react,toolformer,theagentcompany}.
Those ordinary tool observations may contain sensitive data such as absolute paths (e.g., \path{/Users/alice/acme/legal/layoff_bob.pdf}), internal hostnames, branch names, entity names, shell commands, environment variables, and credentials~\cite{panfilov2026stealingreasoningtracesproprietary}. 
Since agents append tool outputs to future turns, these records can become part of the model-facing transcript
even when the user did not explicitly type them.

Unfortunately, once sensitive information enters provider-bound transcripts, they may leave the local trust boundary and be exposed to model providers or, in some cases, other users~\cite{panfilov2026stealingreasoningtracesproprietary}. Consumer AI providers may use user content to improve or train models unless users opt out~\citep{OpenAIModelTraining,DeepSeekPrivacyPolicy}.
Even when model training is disabled, provider-bound transcripts may still
be retained, logged, or cached by serving systems, and operational failures
can expose cross-user metadata or content: for example, a ChatGPT glitch
allowed some users to see titles from other users' chat histories
~\citep{OpenAIChatHistoryBug}. Provider-side inference systems also cache
prompts and prefixes to improve serving efficiency
~\citep{PagedAttention,sglang,ChunkAttention,RAGCache}. 
We need a local boundary that can protect sensitive data while allowing accurate reasoning.

\begin{figure}[t]
    \centering
    \includegraphics[width=0.9\linewidth]{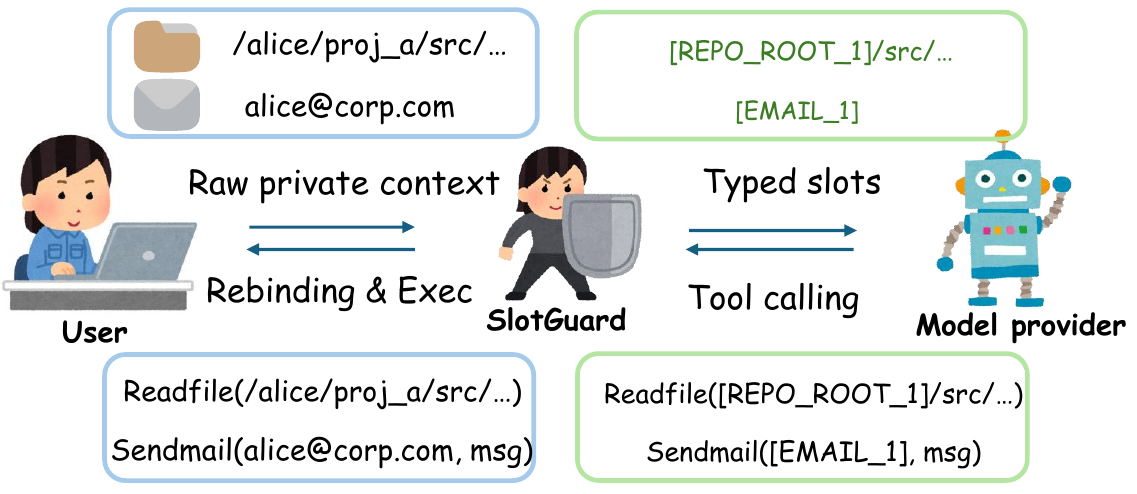}
    \caption{\ourmethod{} sits between the local agent runtime and the upstream provider. Local bindings are replaced with typed placeholders before transcript submission and rebound to raw values inside the trusted local runtime before execution.}
    \label{fig:overview}
\end{figure}

\paragraph{Motivating example}
A local audit of agent session logs can reveal a credential as part of ordinary tool output.
A shell debugging command
prints the full command line of a running experiment with \path{ps}.
The command includes an environment-variable fallback of the form
\path{AZURE_OPENAI_KEY="${AZURE_OPENAI_KEY:-<raw-key>}"}.
Once appended to the agent transcript, the credential-shaped value becomes
provider-bound context and persists across later turns. This failure
mode is easy to miss with assignment-only or vendor-specific scanners:
the value appears inside shell parameter expansion, embedded in a process
listing, rather than as a standalone \path{api_key=<value>} assignment.
It also illustrates why the boundary must interpose on tool outputs and
runtime observations, not only on user-authored prompts.

Our goal is to \textit{hide local secrets and private bindings from model providers while keeping agent transcripts useful}. 
Our system, called \textbf{\ourmethod{}}, transparently sits between agent runtime and upstream models to perform the following operations.
Before transcript submission, \ourmethod{} turns sensitive local values into stable, typed references that a model can still reason about. 
Given responses from the model (before passing them to the agent), \ourmethod{}
resolves those references back to the original values inside the trusted runtime. Ideally, the model should still be able to understand that a value is a Python file, a repository path, a database URL, an API token, or a project-specific handle, without their raw original values.

Intelligent privacy protection mechanisms must distinguish different types of sensitive information.
LLM agent transcripts can leak sensitive data along two surfaces. 
\textbf{\textit{Structural bindings}} identify a user, organization, or execution environment without necessarily granting access: paths, workspace roots, hostnames, branch names, emails, and sensitive entities.
A path may be sensitive because it reveals its owner, customer, project, or semantic content.
\textbf{\textit{Secret values}} are credential-shaped strings that may grant permissions if disclosed, such as API keys, tokens, passwords, DB credentials, and signing material. This distinction is important because these surfaces fail differently: credential scanners can miss semantically sensitive filenames or entity names, while path sanitizers can miss tokens embedded in URLs, shell commands, process listings, or configuration snippets.

Na\"{\i}ve redaction is limited: 
it hides values by destroying the structure that agents need to act. Models use path syntax, file extensions, relative suffixes, URI grammar, and credential-shaped strings as execution cues. Replacing a path with \path{PATH} or a secret with \path{API_KEY} may remove the sensitive value, but it also erases the type, location, and syntactic context needed to plan valid tool calls. The result is a brittle trade-off: strong redaction protects privacy but disrupts workflow compatibility, whereas weak redaction preserves utility but leaves provider-bound transcripts exposed.

To address the issue,
\ourmethod{} establishes a compatibility-preserving local boundary. 
Specifically, \ourmethod{} (1) abstracts structural bindings as typed local slots and (2) substitutes secret values with format-preserving synthetic tokens. The model, therefore, sees stable, executable-looking handles rather than raw private values. Across turns, SlotGuard maintains a session-scoped slot table and a lightweight semantic entity graph that conservatively links protected values to embedded occurrences, derived
forms, and later keyword-based references. When the model requests tool execution, SlotGuard resolves registered handles only inside the trusted runtime and rejects attempts to spoof, traverse, dereference, or otherwise misuse them.
SlotGuard combines best-effort sensitive-span detection with guarded local rebinding; it does not guarantee detection of every sensitive value. It can interpose at the agent harness, where prompts, tool calls, and tool outputs enter the transcript; at a filesystem-facing path layer, where raw paths are virtualized at the source; or at a model-gateway layer such as LiteLLM, which provides a provider-bound last line of defense. Our primary design point is the harness, because it enables both transcript rewriting before provider submission and guarded rebinding before local execution.

We make the following contributions.
\begin{enumerate}[leftmargin=1.2em,itemsep=0pt,topsep=0pt]
  \item We identify structural bindings and secret values as two
  orthogonal privacy surfaces in provider-bound LLM agent transcripts.
  
  \item We introduce \ourmethod{}, a compatibility-preserving local boundary
  that combines typed structural abstraction, format-preserving synthetic
  secret substitution, cross-turn tracking, and guarded local rebinding.
  While many of these techniques were studied individually (e.g., typed placeholders, format-preserving substitution, information-flow tracking), the key contribution is their systematic integration into a single transcript boundary, guided by the observation that preserving execution-relevant structure---path shape, file extensions, credential syntax---is critical for downstream agent performance.

  \item We evaluate \ourmethod{} as an enforceable boundary, measuring
  privacy leakage, workflow compatibility, task success, model
  generalization, and runtime overhead across synthetic privacy corpora, controlled workflow probes, and a TheAgentCompany-derived replay set.
\end{enumerate}

The rest of this paper is organized as follows.
We formalize our threat model in \cref{sec:threat}.
Then, we present the \ourmethod{} design in \cref{sec:design}.
Finally, we evaluate \ourmethod{} in \cref{sec:evaluation}.

\section{Threat Model}\label{sec:threat}

We consider an honest-but-curious upstream provider: inference is correct,
but provider-bound prompts, tool outputs, and transcripts may be retained.
The local runtime, tools, files, and SlotGuard boundary are trusted.
SlotGuard protects the local-to-provider boundary, primarily inside the
agent harness before prompts, tool calls, shell outputs, process listings,
and file contents are appended to the transcript. It protects two value
classes: \emph{structural bindings}, such as paths, workspace roots,
branches, hostnames, emails, and sensitive entity names that identify a
user, organization, project, or environment; and \emph{secret values}, such
as API keys, tokens, passwords, database credentials, and signing material.
The goal is compatibility-preserving privacy: remove raw private values
while preserving workflow signals needed for valid tool use, including
path shape, file type, stable handles, parseable commands, and
format-valid credential surrogates. We measure literal leakage,
categorical leakage (i.e., credential type revealed through replacement-token labels\footnote{%
A placeholder \texttt{\_\_API\_KEY\_abc\_\_} discloses that the redacted value is an API key even when the value itself is withheld
}
),
reconstructive leakage, workflow compatibility, and task success.
These are empirical metrics on annotated workloads, not a guarantee of
full anonymization. Preserved format, suffixes, co-occurrence, and surrounding
context can reveal information. In particular, zero explicit category-label
leakage does not imply that credential type cannot be inferred.

\section{\ourmethod Design}
\label{sec:design}

\paragraph{Boundary and slot identification}
SlotGuard is a local boundary layer between the agent runtime and the
upstream provider. Its primary deployment point is the agent harness, before
prompts, tool calls, tool outputs, shell stdout/stderr, process listings,
and file contents are appended to provider-bound transcripts. This placement
covers failure modes such as credentials printed by \path{ps} or shell
debugging commands. SlotGuard can also be deployed at a filesystem-facing
path layer, where raw paths are virtualized at the source, or at a
model-gateway layer such as LiteLLM, which provides a provider-bound last
line of defense across model backends.

SlotGuard identifies protected spans using a two-tier detector. The
synchronous tier is a heuristic engine for high-precision cases: known
workspace roots, path parsers, emails, hostnames, branch names,
environment-variable assignments, shell parameter expansions, credential
flags, authorization headers, URI credentials, and provider-specific token
patterns. The asynchronous tier optionally uses a small local model for
semantic cases that are difficult to encode as rules, such as sensitive
filenames, entity names, customer names, or project names. The local model
does not rewrite transcripts directly. It proposes a line number, verbatim
span, and slot type; SlotGuard admits the proposal only if the span appears
in the local input and the slot type is allowed for the current task.
This gate validates proposed edits, not detection completeness; sensitive
spans missed by both tiers can remain visible. The microsecond rewrite
measurements exclude this optional semantic pass.

\paragraph{Compatibility-preserving rewriting}
SlotGuard follows a compatibility-preserving privacy principle: remove raw
private values from provider-bound transcripts while preserving the
workflow-level signals agents rely on in unprotected execution. Both
structural bindings and secrets are stored in a session-scoped slot table
$T:V\rightarrow H$, where $V$ is the set of protected raw values and $H$ is
the upstream-visible handle space.

For structural bindings, a session-keyed identifier is computed as
\[
u_s(v)=\mathrm{HMAC}_{k_s}(\tau(v)\parallel v),
\]
where $k_s$ is a per-session secret and $\tau(v)$ is the slot type.
Identifiers are stable within a session; fresh session keys prevent direct
cross-session matching of the keyed identifiers, but do not hide linkage
through retained context or suffixes. We use readable numbered slot
notation in examples, rather than displaying keyed identifiers.
Structural slots include scope roots such as \path{[REPO_ROOT_1]},
identity handles such as \path{[EMAIL_1]}, and sensitive-content handles
such as \path{[SENSITIVE_DOC_1]}. For paths, SlotGuard performs
suffix-aware abstraction: it replaces the longest trusted root, suppresses
sensitive middle components, and preserves a bounded task-relevant suffix:
\path{/Users/alice/acme/legal/layoff_bob.pdf}
$\Rightarrow$
\path{[REPO_ROOT_1]/legal/[SENSITIVE_DOC_1].pdf}.
The result remains path-shaped and preserves directory role and file type,
while removing user, organization, and sensitive document identifiers.

For secret values, SlotGuard uses format-preserving synthetic substitution
(FPS) rather than generic labels. Generic replacements such as
\path{API_KEY} can make the model infer that a credential is missing or
produce commands that no longer match the surrounding grammar. FPS instead
samples a synthetic value from the secret's format class and stores it in
$T$ for reuse at later occurrences. The aim is to retain URI grammar,
credential syntax, and shell configuration cues without exposing the raw
secret. More precisely, let $m(v)$ denote all deliberately preserved format
metadata, including any retained length, prefix, and character alphabet.
Under ideal independent sampling,
\(
z_v\sim\mathrm{Uniform}(\mathcal{F}_{m(v)}),
\)
where $\mathcal{F}_{m(v)}$ is the corresponding format class. The sampled
content is then independent of the secret conditional on $m(v)$.
This is a conditional property of the surrogate-generation rule, not an
end-to-end transcript privacy guarantee: format, repeated occurrences,
and surrounding context can still reveal information.

SlotGuard maintains a session-scoped semantic entity graph (SEG) to extend
protection beyond exact string matches. Nodes are protected values or their
surrogates; edges encode coreference, embedding, derivation, and indirect
reference. 
The SEG extends protection beyond exact string matches by recording conservative links between protected values and observed variants, including embedded occurrences, simple derived forms, and later keyword-based references. It is not intended to provide full semantic understanding; instead, it provides a lightweight session memory for common agent-transcript leakage patterns.
In its current form, the SEG handles keyword- and pattern-based links reliably but does not attempt harder indirect references such as paraphrased coreferences or implicit reasoning chains. We view the current SEG as a conservative first step; extending it with richer coreference resolution or lightweight local models is a natural direction for future work.

\paragraph{Guarded rebinding}
The upstream model only sees typed slots and FPS tokens. Raw values are
restored only inside the trusted runtime and only for validated local
execution. For structural bindings, \sysname checks that a handle exists
in the session table, has the expected type, and rebinds within the allowed
scope. These checks target unknown handles, spoofed segments,
out-of-scope paths, and shell metacharacter injection; the current
prototype's known absolute-path edge case is reported in
\cref{app:stress}. For secrets, an FPS token is resolved to the raw credential only when a local tool call requires it, and the raw value is injected into the process, environment variable, request field, or connection string without being written back to the provider-bound transcript. Fabricated credential-shaped strings that are not registered in the slot table are treated as untrusted literals.

\begin{figure}[t]
\centering
\resizebox{\linewidth}{!}{%
\begin{tikzpicture}[
    font=\scriptsize,
    >=Latex,
    box/.style={
        draw,
        rounded corners,
        fill=white,
        minimum width=2.55cm,
        minimum height=0.85cm,
        align=center
    },
    group/.style={
        draw,
        rounded corners,
        inner sep=10pt
    },
    flow/.style={
        -Latex,
        thick
    },
    lab/.style={
        font=\scriptsize,
        align=center,
        inner sep=1pt
    }
]

\node[box] (agent) at (0,0) {Agent\\Runtime};
\node[box] (slotguard) at (3.3,0) {SlotGuard};
\node[box] (provider) at (7.25,0) {Upstream\\Model Provider};

\begin{pgfonlayer}{background}
  \node[group, fit=(agent)(slotguard), fill=blue!4] (localbox) {};
  \node[group, fit=(provider), fill=red!4] (remotebox) {};
\end{pgfonlayer}

\node[font=\bfseries\small, above=6pt of localbox.north] {Local};
\node[font=\bfseries\small, above=6pt of remotebox.north] {Remote};

\node[font=\scriptsize] at ($(slotguard.north)+(0,0.43)$) {local boundary};

\draw[flow] ([yshift=6pt]agent.east) --
    ([yshift=6pt]slotguard.west)
    node[midway, above=4pt, lab] {tool I/O};

\draw[flow] ([yshift=-6pt]slotguard.west) --
    ([yshift=-6pt]agent.east)
    node[midway, below=4pt, lab] {guarded rebinding};

\draw[flow] ([yshift=7pt]slotguard.east) --
    ([yshift=7pt]provider.west)
    node[midway, above=3pt, lab] {sanitized transcript};

\draw[flow] ([yshift=-7pt]provider.west) --
    ([yshift=-7pt]slotguard.east)
    node[midway, below=3pt, lab] {response};

\end{tikzpicture}%
}

\caption{
\ourmethod{} runs locally between the agent runtime and the upstream model
provider. Sanitized transcripts go upstream; raw values are rebound only
inside the trusted runtime.
}
\label{fig:slotguard-architecture}
\end{figure}
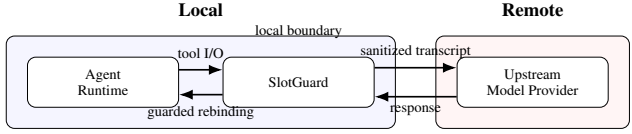

\section{Evaluation}
\label{sec:evaluation}

We evaluate SlotGuard as a local transcript boundary: it should remove
provider-visible bindings while preserving the workflow cues needed for
local agent operations.\footnote{Artifacts: \codebaseurl} We
study four questions: structural leakage (RQ1), credential leakage (RQ2),
workflow compatibility (RQ3), and hot-path practicality (RQ4).

\paragraph{Setup}
We use three workloads: two synthetic privacy corpora (a 30-session structural corpus with 9{,}229 paths and a 200-session credential corpus with 852 planted values), synthetic workflow probes (a 12-task local probe and a 200-task multi-model probe), {and a derived replay set, TAC-lite, from The Agent Company with 33 task families, 40 grounded handles, and 160 replay tasks~\cite{theagentcompany}}. We compare \emph{Raw}, \emph{Generic redaction}, a restore-aware \emph{VibeGuard-style typed-placeholder baseline} derived from \path{default.vgrules}\footnote{\url{https://raw.githubusercontent.com/inkdust2021/vgrules/refs/heads/main/default.vgrules}}, and \emph{Full \ourmethod{}}. Component ablations and
expanded setup details are deferred to~\cref{app:eval-details}.

\subsection{RQ1--RQ2: Does SlotGuard remove private values?}
Table~\ref{tab:privacy-summary} summarizes the core privacy result. Across
the structural corpus, raw transcripts expose 20{,}814 annotated sensitive
characters, while \ourmethod{} removes all annotated structural
leakage in this corpus. The rewrite still preserves execution cues: 93.6\% of rewritten
paths keep the original extension and 100.0\% keep at least one safe suffix
component. Across the credential corpus, typed placeholders and FPS-only (format-preserving synthetic substitution in \cref{sec:design})
still leak 23.5\% of planted credentials because split and embedded
exposures remain reconstructible, whereas the full version with SEG (semantic entity graph in \cref{sec:design}) leaks 0/200 sessions
and 0/852 planted values. Detailed breakdowns, including format-validity
results, appear in~\cref{app:eval-details}.

\begin{table}[t]
\centering
\small
\caption{Key privacy results. Structural rows compare against raw
transcripts; credential rows compare against the strongest 
baseline.}
\label{tab:privacy-summary}
\begin{tabular}{lrr}
\toprule
Metric & Baseline & SlotGuard \\
\midrule
\multicolumn{3}{l}{\textit{Structural privacy / path compatibility}} \\
Sensitive chars visible upstream & 20{,}814 & 0 \\
Structural leakage & 100.0\% & 0.0\% \\
Paths retaining file extension & -- & 93.6\% \\
Paths retaining safe suffix & -- & 100.0\% \\
\addlinespace[2pt]
\multicolumn{3}{l}{\textit{Credential leakage / secret exposure}} \\
Credential session leak & 100.0\% & 0.0\% \\
Credential leak rate & 23.5\% & 0.0\% \\
\bottomrule
\end{tabular}
\end{table}

\subsection{RQ3: Does \ourmethod{} retain performance?}

The local probe shows the basic pattern: generic redaction
collapses to 16.7\% success, VibeGuard-style~\cite{VibeGuard} placeholders recover some signal at 66.7\%, and full \ourmethod{} reaches 83.3\% with no invalid handles. We keep the local probe and ablations in \Cref{app:workflow-details}.
The larger 200-task probe across four upstream models shows the same
ordering (\Cref{tab:rq3-modelgen}). Generic redaction collapses to
2.5\% task success for every model. VibeGuard-style typed placeholders are
better than opaque replacement, but still leave a large gap, ranging from
20.5\% to 38.5\%. In contrast, full \ourmethod{} stays close to each
model's raw baseline: it matches the raw score for Kimi and Llama, remains
within 2.0 points for \texttt{gpt-5.4}, and within 10.5 points for
DeepSeek. The main effect is therefore not merely that ``typed placeholders
help''; it is that path-shaped and format-preserving rewrites preserve far
more execution-relevant structure than proxy-style placeholders.

\begin{table}[t]
\centering
\footnotesize
\caption{Model generalization on 200 shared workflow tasks.}
\label{tab:rq3-modelgen}
\begin{tabular}{p{0.28\linewidth}rrrr}
\toprule
Model & Raw & Generic & VG-style &  \ourmethod{} \\
\midrule
DeepSeek-V3.2 & 100.0\% & 2.5\% & 38.5\% & 89.5\% \\
gpt-5.4 & 100.0\% & 2.5\% & 20.5\% & 98.0\% \\
Kimi-K2.6 & 100.0\% & 2.5\% & 21.5\% & 100.0\% \\
Llama-3.3-70B & 93.5\% & 2.5\% & 31.5\% & 93.5\% \\
\bottomrule
\end{tabular}
\end{table}

{We also evaluate the replay set, TAC-lite. TAC-lite extracts evaluator-grounded repo-relative handles and instantiates local replay tasks without TAC's Dockerized service stack. The same ordering reappears: generic redaction falls to 0.0\% across all four models, so we omit it. VibeGuard-style placeholders reach only 26.9--36.9\%, and full \ourmethod{} matches the raw score for \texttt{gpt-5.4}, Kimi, and Llama while remaining within 7.5 points on DeepSeek. \Cref{app:tac-lite} describes the extraction procedure; \cref{tab:rq3-tac-lite} reports replay handle-selection success, not end-to-end TAC task completion.}

While our evaluation employs diverse data such as (1) synthetic privacy corpora, (2) controlled workflow probes, and (3) a replay-style benchmark derived from TheAgentCompany,
they may not fully capture the diversity of real-world enterprise agent traces. 
Validation on production-like environments with naturally occurring sensitive data remains an important direction for future work.

\begin{table}[t]
\centering
\small
\caption{TAC-lite replay handle-selection success.}
\label{tab:rq3-tac-lite}
\begin{tabular}{p{0.34\linewidth}rrrr}
\toprule
Model & Raw & VG-style & \ourmethod{} \\
\midrule
DeepSeek-V3.2 & 95.0\% & 34.4\% & 87.5\% \\
gpt-5.4 & 92.5\%  & 36.9\% & 92.5\% \\
Kimi-K2.6 & 95.0\% & 35.6\% & 95.0\% \\
Llama-3.3-70B & 90.0\% & 26.9\% & 90.0\% \\
\bottomrule
\end{tabular}
\end{table}

\subsection{RQ4: Is \ourmethod{} practical on the hot path?}

The deterministic rewrite has low overhead in the measured microbenchmarks. Rewriting a path takes 1.690~$\mu$s median, mapping it back takes 0.398~$\mu$s, and sanitizing a full agent turn takes 14.424~$\mu$s; even a 6\,KB turn stays below 100~$\mu$s median. 
The optional local semantic detector is much slower (742~ms warm median), so SlotGuard does not rely on it for the fast path. Instead, the detector only suggests candidate sensitive spans, and SlotGuard verifies each suggestion before applying a rewrite. On a held-out sensitive-filename benchmark, this optional detector improves $F_1$ from 0.571 to 0.909. All 9{,}229 rewritten paths round-trip correctly, and all 64 crafted invalid rebinding attempts are rejected. In the mixed hard-case suite, 23 of 24 inputs are rejected or sanitized, with one known absolute-path miss. Additional details are in
\Cref{app:eval-details}.

\section{Conclusion}

\ourmethod is a local transcript boundary for LLM agents. It removes
structural bindings with typed, suffix-aware slots and protects credentials
with format-preserving synthetic values. Across our workloads, SlotGuard
eliminates annotated structural leakage and credential leakage while
remaining close to raw-transcript task success on synthetic probes and
raw-transcript handle-selection success on TheAgentCompany-derived replay tasks. SlotGuard
is not full anonymization, but a step toward local capability boundaries that
mediate both transcript privacy and tool access.

Several directions remain open. First, our evaluation uses synthetic and replay-style workloads with planted ground truth; validating SlotGuard on production agent traces with naturally occurring sensitive data would strengthen the practical evaluation. A small real-world deployment case study is a natural next step. Second, we do not measure adversarial reconstruction by a capable attacker who actively attempts to recover protected values from residual context; such an evaluation would complement the current literal- and reconstructive-leakage proxies. Third, the semantic entity graph is an early implementation that tracks entities through keyword- and pattern-based links; standalone verification on harder indirect-reference benchmarks and extending it with richer coreference resolution would make the component more robust. Finally, comparison against production secret scanners, enterprise DLP systems, or LLM-specific guardrail products would better situate SlotGuard's detection coverage relative to deployed baselines.

\section*{Impact Statement}

This work introduces a privacy boundary for tool-using LLM agents. Its
intended benefit is to reduce unnecessary exposure of local identifiers and
credentials to upstream model providers while preserving agent utility.
\ourmethod{} is not a full anonymization system: it cannot prevent semantic
inference from non-literal context, protect a compromised local runtime, or
stop prompt-injection-driven exfiltration. 
The current evaluation does not measure adversarial reconstruction by a capable attacker who actively attempts to recover protected values from residual transcript context; such an evaluation would further stress-test the boundary's practical guarantees. 
It should therefore be deployed
as one layer in a broader privacy posture. We do not identify a specific
dual-use risk; the mechanism reduces provider-bound information flow and
runs under user control.

\section*{Acknowledgements}

This work was supported in part by the National Science Foundation under grants \#2312561 and \#2440498. This work used Delta and DeltaAI at the National Center for Supercomputing Applications (NCSA) through allocation CIS240661 from the Advanced Cyberinfrastructure Coordination Ecosystem: Services \& Support (ACCESS) program, which is supported by U.S.\ National Science Foundation grants \#2138259, \#2138286, \#2138307, \#2137603, and \#2138296. This work was also supported by the IBM-Illinois Discovery Accelerator Institute.

\bibliography{main}
\bibliographystyle{icml2026}

\newpage
\appendix
\onecolumn

\section{Related Work}
\paragraph{Plugin-level redaction}
Open-VibeGuard~\citep{OpenCodeVibeGuard}
applies harness-level redaction rules, such as an SSN rule matching
three, two, and four digits separated by hyphens, and replaces matched
values with stable category-hash placeholders such as
\path{__VG_SSN_xxxxxxxxxxxx__}. This is a
practical, lightweight defense for matched spans, but rule-based placeholder
redaction is local and structurally brittle. It can over-redact benign
lookalikes that share the same surface pattern, while missing sensitive
values embedded in shell expansions, URIs, process listings, and multi-line
configuration snippets, derived forms, or later cross-turn references. In
agent workflows, placeholders may also be mistaken for missing values or
copied into commands, URIs, and configuration snippets in forms that no
longer match the expected grammar. \ourmethod{} addresses these gaps with a
session-scoped slot table and SEG, suffix-aware structural slots,
format-preserving synthetic secrets, and guarded local rebinding.

\paragraph{Network-level proxies}
Network-boundary redaction systems interpose below the application, for
example as an HTTPS proxy with regex or NER-based rewriting
~\citep{VibeGuard}. This placement can cover multiple clients without
agent-specific integration, but it only sees serialized provider requests.
It does not naturally expose agent structure such as tool calls, tool
outputs, shell observations, local scopes, or execution intent. As a result,
network proxies are well suited as last-line defenses, but cannot easily
perform scope-aware path validation or rebind protected handles back into
trusted local tool execution. They may also require trusted-CA installation,
changing the user's network trust assumptions.

\paragraph{PII detection and anonymization}
General-purpose PII systems such as Microsoft Presidio~\citep{presidio}
detect entities such as names, emails, phone numbers, and identifiers, and
replace them with labels or anonymized values. These systems target text
anonymization, whereas agent transcripts require compatibility-preserving
rewriting. A replacement must hide the raw value while preserving the cues
models rely on for tool use: path shape, file extensions, relative suffixes,
URI structure, header grammar, and credential-shaped syntax. SlotGuard
therefore treats transcript rewriting as part of an execution boundary, not
only as entity anonymization.

\paragraph{Information flow control}
SlotGuard is conceptually related to dynamic taint analysis and
information-flow control, which track sensitive values as they propagate
through computation~\citep{clause2007dytan}. Our protected sink is the
provider-bound transcript of an LLM agent. The semantic entity graph plays a
taint-like role by linking raw values, rewritten handles, embedded
occurrences, derived forms, and later indirect references. The policy goal
also follows contextual integrity~\citep{nissenbaum2004privacy}: local values may
be appropriate inside the trusted runtime, but inappropriate once copied
into a remote model-provider transcript.

\section{Motivating Transcript Leakage Example}
\label{app:motivating-leak}

During a local audit of agent session logs, we found that credentials can
enter provider-bound transcripts through ordinary tool output rather than
through user-authored prompts. One example came from a shell debugging
command that printed a process listing. The resulting transcript contained
a command line with an environment-variable fallback:

\begin{slotlisting}[Example of session snippet]
AZURE_OPENAI_KEY="${AZURE_OPENAI_KEY:-<AZURE_OPENAI_KEY_VALUE>}" uv run python experiment.py --data_path ... --resume
\end{slotlisting}

This pattern is easy to miss with vendor-specific or assignment-only
secret scanners. The secret is not presented as a standalone API key, nor
as a simple assignment of the form \texttt{api\_key=<value>}. Instead, it
appears inside shell parameter expansion, embedded in a process listing,
and then becomes part of the agent transcript when the tool output is
appended to the conversation.

SlotGuard treats shell commands, process listings, environment assignments,
and parameter-expansion fallbacks as credential-bearing contexts. In this
case, the raw fallback value would be replaced by a format-preserving
synthetic surrogate before transcript submission:

\begin{slotlisting}[Processed by SlotGuard (schematic)]
AZURE_OPENAI_KEY="${AZURE_OPENAI_KEY:-<FPS_AZURE_OPENAI_KEY>}" uv run python experiment.py --data_path ... --resume
\end{slotlisting}

The angle-bracketed strings above are explanatory markers, not literal
SlotGuard output. In actual rewriting, the fallback contains a synthetic
value in the credential's format class, rather than the displayed
\texttt{<FPS\_AZURE\_OPENAI\_KEY>} label. This preserves shell syntax and
configuration cues without transmitting the raw fallback credential;
format and surrounding context remain visible. During validated
local execution, the FPS token is resolved through the session slot table
inside the trusted runtime.

\section{Additional Evaluation Details}
\label{app:eval-details}

\paragraph{Expanded setup}
The structural corpus contains 30 randomized repository-like sessions with
9{,}229 paths and transcript fragments containing user-home prefixes,
workspace roots, repository roots, emails, internal hosts, and sensitive
document names. The credential corpus contains 200 sessions with 852 planted
credentials spanning direct, embedded, split, derived, and cross-turn
contexts. The workflow suite contains a 12-task local probe against
\texttt{qwen2.5-coder:1.5b} via an OpenAI-compatible Ollama endpoint and a
200-task shared probe across four Azure-hosted models. The
VibeGuard-style baseline is derived from \path{default.vgrules} categories such
as \path{EMAIL}, \path{API_KEY}, \path{BEARER_TOKEN}, \path{JWT},
\path{AWS_KEY}, and \path{PRIVATE_KEY}, and rewrites matches to restore-aware
placeholders of the form \path{__VG_<CATEGORY>_<hash>__}
\cite{VibeGuard,OpenCodeVibeGuard}.

\subsection{Detailed Privacy Results}
\label{app:privacy-details}

\begin{table}[t]
\centering
\small
\caption{Structural privacy and path-compatibility results.}
\label{tab:rq1-structural}
\begin{tabular}{lrr}
\toprule
Metric & Baseline & SlotGuard \\
\midrule
Sensitive chars visible upstream & 20{,}814 & 0 \\
Structural char leakage & 100.0\% & 0.0\% \\
Paths retaining file extension & -- & 93.6\% \\
Paths retaining $\geq 1$ safe suffix component & -- & 100.0\% \\
Fully opaque paths & -- & 0.0\% \\
\bottomrule
\end{tabular}
\end{table}

\begin{table}[t]
\centering
\small
\caption{Credential leakage on 200 sessions / 852 planted values.}
\label{tab:rq2-credential}
\begin{tabular}{lrrr}
\toprule
Method & Session leak & Credential leak & Explicit category label \\
\midrule
No protection & 100.0\% & 100.0\% & N/A \\
Typed placeholder baseline & 100.0\% & 23.5\% & 100.0\% \\
FPS only & 100.0\% & 23.5\% & 0.0\% \\
\ourmethod{} (FPS+SEG) & 0.0\% & 0.0\% & 0.0\% \\
\bottomrule
\end{tabular}
\end{table}

Session leakage is the fraction of the 200 sessions with at least one
exposed planted credential; credential leakage is measured over the 852
planted values, including reconstructible split exposures. The explicit
category-label metric concerns type names in replacement labels only,
not credential-type inference from format or context. It is not applicable
to unmodified transcripts. Structural character leakage is measured over
the 20{,}814 annotated sensitive characters in the structural corpus.

The split-exposure cases highlight the main limitation of non-graph methods:
all 200 split cases leak under both typed placeholders and FPS-only, whereas
FPS+SEG eliminates this leakage. We also evaluate whether generated
surrogates preserve the grammar of their original contexts. Across API keys,
OAuth-style tokens, private-key markers, and internal hostnames, parse
success ranges from 83.3\% to 100.0\%; the remaining failures are concentrated
in deliberately awkward PII and multiline edge cases, rather than in the
credential classes used by the workflow probes.

We emphasize that the current SEG is still an early implementation. It
tracks protected entities through keyword- and pattern-based links rather
than full semantic understanding, so it should be viewed as a conservative
first step toward semantic transcript memory rather than a complete solution
to indirect leakage.
A dedicated standalone evaluation of the SEG---measuring recall on diverse indirect-reference patterns such as paraphrased mentions, implicit coreference, and multi-hop derivation chains---would help quantify its coverage gaps and guide improvements. We consider this an important direction for strengthening the component.

\subsection{Workflow Ablations}
\label{app:workflow-details}

\begin{table}[t]
\centering
\small
\caption{Workflow compatibility on the local 12-task probe with ablations.}
\label{tab:rq3-local-full}
\begin{tabular}{p{0.36\linewidth}rrrr}
\toprule
Condition & Task success & Exec.-ready & Invalid handles & Avg. latency \\
\midrule
Raw & 100.0\% & 100.0\% & 0.0\% & 1551 ms \\
Generic redaction & 16.7\% & 0.0\% & 83.3\% & 637 ms \\
VibeGuard-style placeholder & 66.7\% & 33.3\% & 0.0\% & 724 ms \\
No FPS & 50.0\% & 66.7\% & 0.0\% & 718 ms \\
No suffix preservation & 50.0\% & 0.0\% & 0.0\% & 805 ms \\
Full \ourmethod{} & 83.3\% & 66.7\% & 0.0\% & 719 ms \\
\bottomrule
\end{tabular}
\end{table}

The ablations separate the two design contributions. Removing FPS preserves
some path utility but weakens secret handling, while removing suffix
preservation retains the measured privacy protection but makes structural responses much
less execution-ready. SlotGuard needs both components to preserve the
workflow signal that raw transcripts provide.

\subsection{TAC-lite Grounded-Handle Replay}
\label{app:tac-lite}

To reduce dependence on fully synthetic task wording, we derive a
lightweight replay benchmark from The Agent Company. We do \emph{not} run
the original TAC Dockerized multi-service stack. Instead, we scan
\path{task.md} and evaluator code for grounded local file handles (e.g.,
\path{open(...)} targets, evaluator working directories, and destination
paths), retain repo-centric tasks that resolve to local workspace files, and
instantiate each grounded handle under four randomized local workspace roots.
This yields 33 source task families, 40 grounded handles, and 160 replay
tasks. Because every item is structural, the main metrics are exact handle
selection, execution-ready rate, tool-valid rate, and invalid-handle rate.
The four root variants share source handles and task families; the 160
items should not be interpreted as 160 independent real-world tasks or as
end-to-end executions of the original TAC benchmark.

\begin{table}[t]
\centering
\small
\caption{TAC-lite handle-selection success on 160 evaluator-grounded replay
items derived from The Agent Company; not end-to-end TAC task completion.}
\label{tab:tac-lite}
\begin{tabular}{p{0.30\linewidth}rrrr}
\toprule
Model & Raw & Generic & VG-style & Full \ourmethod{} \\
\midrule
DeepSeek-V3.2 & 95.0\% & 0.0\% & 34.4\% & 87.5\% \\
gpt-5.4 & 92.5\% & 0.0\% & 36.9\% & 92.5\% \\
Kimi-K2.6 & 95.0\% & 0.0\% & 35.6\% & 95.0\% \\
Llama-3.3-70B & 90.0\% & 0.0\% & 26.9\% & 90.0\% \\
\bottomrule
\end{tabular}
\end{table}

The TAC-lite results mirror the synthetic workflow probe. Generic redaction
collapses to 0.0\% handle-selection success for every model because the returned handles
are typically unrebindable or structurally invalid. VibeGuard-style
placeholders recover some semantics, but only reach 26.9--36.9\% exact
handle success. Full \ourmethod{} matches the raw score on
\texttt{gpt-5.4}, Kimi, and Llama, and remains within 7.5 points on
DeepSeek. Tool-valid rates for full \ourmethod{} remain in the
92.5--97.5\% range, close to raw, whereas generic redaction induces
invalid-handle rates between 70.0\% and 100.0\%.

\subsection{Model-Dependent End-to-End Latency}
\label{app:latency-details}

End-to-end latency depends on both the model and the transcript
representation. Our accounting includes request backoff and JSON-parse
retries, whereas deterministic local rewriting is measured separately at
microsecond scale. One possible explanation for representation-dependent
latency is that placeholders such as \path{__VG_API_KEY_<hash>__} and
\path{__VG_PATH_<hash>__} provide weaker structural cues and alter retry
frequency. The aggregate latency measurements do not isolate this causal
mechanism.

\begin{table}[t]
\centering
\small
\caption{Average end-to-end latency on the 200-task model-generalization
probe.}
\label{tab:vg-latency}
\begin{tabular}{lrr}
\toprule
Model & VG-style & Full \ourmethod{} \\
\midrule
DeepSeek-V3.2 & 1283 ms & 4701 ms \\
gpt-5.4 & 684 ms & 701 ms \\
Kimi-K2.6 & 26174 ms & 5774 ms \\
Llama-3.3-70B & 570 ms & 856 ms \\
\bottomrule
\end{tabular}
\end{table}

Table~\ref{tab:vg-latency} shows no universal latency advantage.
Full SlotGuard is faster on \texttt{Kimi-K2.6}, but slower on the other
three models, with a small difference on \texttt{gpt-5.4}. These results
should be read alongside the utility differences, rather than as a general
speedup claim. A similar model-dependent pattern appears on TAC-lite: on
\texttt{Kimi-K2.6}, VG-style placeholders average 10.97\,s per task versus
3.16\,s for full SlotGuard, while on \texttt{gpt-5.4} the two conditions
remain close (492\,ms vs.\ 459\,ms).

\subsection{Hot-Path and Semantic-Layer Details}
\label{app:hotpath-details}

The core deterministic path remains negligible relative to provider
inference: \path{abstract_path} has a median cost of 1.690~$\mu$s,
\path{rebind_path} 0.398~$\mu$s, and \path{abstract_text/agent_turn}
14.424~$\mu$s. A larger 6\,KB transcript-sized turn remains under
100~$\mu$s median (96.803~$\mu$s). On a 30-example held-out filename probe,
rules alone achieve precision $=1.00$, recall $=0.40$, and $F_1=0.571$.
Adding a small local CPU-runnable model raises recall to $1.00$ and
$F_1=0.909$. The semantic pass has a cold first-call latency of 919~ms, a
warm median of 742~ms, and a p95 of 929~ms, so we keep it optional and
advisory under a verify-then-replace gate.

\subsection{Hard and Adversarial Cases}
\label{app:stress}

\begin{table}[t]
\centering
\small
\caption{Hard and adversarial cases. ``Rejected'' means input refused with
explicit error; ``sanitized'' means input neutralized. One current
out-of-scope absolute-path edge case remains.}
\label{tab:stress}
\begin{tabular}{lcc}
\toprule
Category & N & Outcome \\
\midrule
\path{..} traversal in rebinding & 6 & 6 rejected \\
Spoofed placeholder not in slot table & 4 & 4 rejected \\
Shell metacharacters (\texttt{; | \& \$ `}) & 5 & 5 rejected \\
Out-of-scope absolute paths & 3 & 2 rejected \\
Mixed encoding / non-UTF-8 fragments & 2 & 2 sanitized \\
Pathological size (path $>$10\,KB) & 2 & 2 sanitized \\
Bracket syntax in legitimate filenames & 2 & 2 sanitized \\
\midrule
\textbf{Total} & \textbf{24} & \textbf{23/24 contained} \\
\bottomrule
\end{tabular}
\end{table}

Across all 9{,}229 rewritten paths, guarded rebinding succeeds on every
round-trip, with zero mismatches and zero unexpected errors. On adversarial
inputs, SlotGuard rejects all 64 crafted invalid rebindings, including
traversal attempts, spoofed placeholders, and shell-metacharacter
injections. This rebinding result is distinct from the mixed 24-case
hard-input suite in \cref{tab:stress}, where 23 of 24 inputs are rejected
or sanitized.
These finite checks do not establish fail-closed behavior for every
adversarial input. The lone hard case-suite miss is a root-owned absolute path
\path{/root/.ssh/id_rsa}, which exposes a remaining out-of-scope
home-prefix edge case in the current prototype rather than a rebinding
failure.

\end{document}